\documentstyle [twocolumn,prb,aps] {revtex}
\begin{document}
\draft
\begin{title}
{Formation and control of electron molecules in artificial atoms:\\
Impurity and magnetic-field effects}
\end{title} 
\author{Constantine Yannouleas and Uzi Landman} 
\address{
School of Physics, Georgia Institute of Technology,
Atlanta, Georgia 30332-0430 }
\date{To appear in Phys. Rev. B {\bf 61} [15 June 2000]}
\maketitle
\begin{abstract}
Interelectron interactions and correlations in quantum dots can lead to
spontaneous symmetry breaking of the self-consistent mean field resulting in 
formation of Wigner molecules. With the use of spin-and-space unrestricted 
Hartree-Fock (sS-UHF) calculations, such symmetry breaking is discussed for 
field-free conditions, as well as under the influence of an external magnetic 
field. Using as paradigms impurity-doped (as well as the limiting case of 
clean) two-electron quantum dots (which are analogs to helium-like atoms), it 
is shown that the interplay between the interelectron repulsion and the 
electronic zero-point kinetic energy leads, for a broad range of impurity 
parameters, to formation of a singlet ground-state electron molecule, 
reminiscent of the molecular picture of doubly-excited helium. Comparative 
analysis of the conditional probability distributions for the sS-UHF and the 
exact solutions for the ground state of two interacting electrons in a clean
parabolic quantum dot reveals that both of them describe formation of an 
electron molecule with similar characteristics. The self-consistent field
associated with the triplet excited state of the two-electron quantum dot 
(clean as well as impurity-doped) exhibits symmetry breaking of the 
Jahn-Teller type, similar to that underlying formation of nonspherical 
open-shell nuclei and metal clusters. Furthermore, impurity and/or 
magnetic-field effects can be used to achieve controlled manipulation of the 
formation and pinning of the discrete orientations of the Wigner molecules. 
Impurity effects are futher illustrated for the case of a quantum dot with 
more than two electrons.
\end{abstract}
~~~~\\
\narrowtext

\section{Introduction}

Two-dimensional (2D) quantum dots (QD's) created at semiconductor interfaces
with refined control of their size, shape and number of electrons 
are often referred \cite{kast,asho,taru} to as ``artificial atoms''. This 
analogy suggests that the physics of electrons in such
man-made nanostructures is closely related to that underlying the
traditional description \cite{cs} of natural atoms (pertaining particularly 
to electronic shells and the Aufbau principle), where the electrons are taken
\cite{hart} to be moving in a spherically averaged effective central mean 
field (CMF). However, using as paradigms impurity-doped (as well as the 
limiting case of clean) two-electron-QD ($2e$ QD) analogs to He-like atoms, 
we show that the interplay between the interelectron repulsion ($Q$) and the
electronic zero-point kinetic energy ($K$) may lead, for a broad range of
impurity parameters,  to spontaneous 
symmetry-breaking (SB) of the self-consistent mean field, resulting at zero 
magnetic field ($B=0$) in formation of a singlet ground-state electron 
molecule. Such SB is beyond the CMF picture and, while
negligible in the ground state of the He atom (whose study was
central to the development of the quantum theory of matter due to the
failure of the Bohr-type models \cite{note111}), it is similar in nature to 
the SB found \cite{note111} in the 1970's in doubly excited He, where formation
of an $e$-He$^{2+}$-$e$ ``triatomic'' molecule has been invoked. Furthermore,
we show that symmetry breaking at $B=0$ of the self-consistent field 
associated with the triplet excited state of the 
$2e$ QD originates from a Jahn-Teller distortion of the CMF, similar to that 
underlying formation of nonspherical open-shell nuclei \cite{bm,naza} and 
metal clusters. \cite{yl1,ekar} 

Along with a
unification of concepts pertaining to spontaneous SB in a variety of finite
fermionic systems (from nuclei, metal clusters and natural atoms, to 2D QD's), 
we demonstrate the ability to control the orientation and to manipulate 
(i.e., to enhance, but also to counteract and even to void) the formation 
of the electron molecules in 2e QD's via 
impurity and/or magnetic-field effects. The ability to control the orientation
of the electron molecule may in principle open new possibilities for 
designing ``on-off'' (i.e., switching between two discrete states) devices, 
which eventually may be employed in applications of QD's as nanoscale logic 
gates \cite{lent} (the effect of impurities on the structure of multi-electron
molecules in QD's with more than two electrons is further illustrated in 
Appendix A).

That electrons in extended media may undergo crystallization 
at low densities, when $Q$ dominates over $K$, has been 
predicted \cite {wign} by Wigner in 1934. Such Wigner crystallization (WC) 
in clean QD's results in formation of electron molecules 
\cite{maks,mk,grab,yl2} [also referred 
to as Wigner molecules (WM's)], which are associated with spontaneous
SB, where the symmetry of the ground state, calculated at the mean-field 
(self-consistent-potential) level, is found to be lower than that of the 
exact hamiltonian describing the system. \cite{ande,thou,fuku}
In clean QD's, formation \cite{lp} of WM's is controlled \cite{yl2,noterw} 
by the parameter $R_W=Q/K$. For a parabolic confinement 
(with frequency $\omega_0$) at $B=0$, 
it is customary to take $Q=e^2/\kappa l_0$ and $K \equiv \hbar \omega_0$,
where $\kappa$ is the dielectric constant and $l_0=(\hbar/m^* \omega_0)^{1/2}$
is the spatial extent of the lowest state's wave function of an electron with 
an effective mass $m^*$; WM's occur \cite{yl2} for $R_W > 1$, corresponding
to much higher electron densities \cite{grab,yl2} than those predicted for 
WC in an infinite 2D medium. \cite{tc}

The many-body hamiltonian for a QD with $N_e$ electrons can be expressed as 
a sum of a single-particle part and the two-particle interelectron 
Coulomb repulsion,
\begin{equation}
{\cal H}=\sum_{i=1}^{N_e} H(i) + 
\sum_{i=1}^{N_e} \sum_{j>i}^{N_e} \frac{e^2}{\kappa r_{ij}}~,
\label{mbh}
\end{equation}
The contributions to the single-particle part are written as
\begin{equation}
H(i)=H_0(i)+H_B(i)+V_I(i)~,
\label{hi}
\end{equation}
and they contain a term describing
the motion of an electron in a 2D parabolic confinement, i.e., 
$H_0 (i) =  { {\bf p}_i^2 }/{2m^*} + m^* \omega_0^2 ( x_i^2 + y_i^2 )/2$,
where $\omega_0$ is the frequency of the 2D isotropic harmonic confining
potential. Magnetic-field effects are included in 
$H_B(i) = [({\bf p}_i - e{\bf A}_i/c)^2 - {\bf p}_i^2 ]/2m^*
+ g^* \mu_B {\bf B} {\bf \cdot} {\bf S}_i/\hbar$,
where the vector potential, ${\bf A}_i = B(-y_i/2,x_i/2,0)$, is taken in the
symmetric gauge, and the last term is the
Zeeman interaction with an effective factor $g^*$, ${\bf S}_i$ is the
electron spin, and $\mu_B$ the Bohr magneton.
To include the effect of (Coulombic) impurities, we added to 
$H(i)$ the term 
$V_I(i) = (e/\kappa) \sum_l {\cal Q}_l/|{\bf r}_i - {\cal R}_l|$ 
($i=1,...,N_e$), where ${\cal Q}_l$ is the charge of the $l$th impurity
located at ${\cal R}_l = (x_l,y_l,d_l)$;
such impurities which in general may be situated out of the
2D plane of the QD (that is $d_l \neq 0$) may correspond to implanted 
atoms (donors or acceptors) or represent a fabricated, controllable 
voltage-gate. 

A clean QD (that is with ${\cal Q}_l=0$ in the above hamiltonian) may 
be regarded as a realization of the Thomson atom, as are 
jellium models of metal clusters, \cite{ekar} where the positive charge is
uniformly distributed; in the $2e$ Thomson QD (TQD), the confinement
to the 2D plane is modeled by a parabolic potential. On the other hand,
a QD analog of the Rutherford model of the atom (RQD), where the positive
charges are grouped in a single nucleus, can be achieved through the
introduction of a central attractive impurity (in addition to the
harmonic confinement).

The presence of many free parameters in Eqs. (\ref{mbh}) and (\ref{hi}) 
results in a plethora of possible case studies generated by varying 
the material dependent parameters ($\kappa$ and $m^*$), the harmonic
confinement $\omega_0$ and the magnetic-field $B$, as well as the
number of impurities $l$, their charges ${\cal Q}_l$, and positions 
${\cal R}_l$. In this paper, we have chosen to discuss here the 
following three representative classes of cases 
at a specific value of $R_W=2.39$, which is sufficientlly high so 
that the electrons form \cite{yl2} a WM in the case of a clean QD,
thus allowing for systematic investigations of the effects of
impurities on the formation, orientational pinning, and structural distortions
of the electron molecules (for the other parameter values used throughout this
paper, see Ref.\ \onlinecite{note11}). The three representative classes, which
we discuss, are: (i) A 2e TQD at $B=0$ in the presence of two off-centered 
impurities situated on both sides of the dot (section II); (ii) A 2e 
RQD (with a single attractive central impurity) in the 
presence of an applied magnetic field (for both the cases of a weak and a 
strong field, section III); and (iii) A QD with eight electrons at $B=0$ and a
single impurity with varying strength, polarity, and location (Appendix A).

For the case of a clean 2e QD, with the hamiltonian given by Eqs.\
(\ref{mbh}) and (\ref{hi}) with $N_e=2$ and $V_I=0$, the exact solution can 
be found quite easily \cite{wagn,pfann,taut1,taut,mol}
owing to the separability of the Schr\"{o}dinger equation in
the center-of-mass and relative coordinates; in the presence of impurities,
separability is lost and finding an exact solution \cite{berk}
becomes significantly more complicated (even for two electrons in the
presence of a central impurity). Therefore, and also in order to expound the
physical principles underlying spontaneous symmetry breaking in QD's, 
we will mostly use in the following (sections II and III and Appendix A) the 
self-consistent spin-and-space unrestricted Hartree-Fock (sS-UHF) method 
\cite{note12} which, unlike the restricted HF (RHF) technique, \cite{so,grh} 
allows for the formation of broken-symmetry states (of lower energy than those
obtained via the RHF). This sS-UHF, which we introduced for studies of QD's in 
Ref.\ \onlinecite{yl2}, employs $N_e$ (where $N_e$ is the number of 
electrons) orbital-dependent, effective (mean-field) potentials and it
differs from the usual \cite{hf} RHF in two ways: (i) it relaxes the 
double-occupancy requirement, namely, it employs different spatial orbitals 
for the two different (i.e., the up and down) spin 
directions [DODS, thus the designation ``spin (s) unresricted''], and (ii) it 
relaxes the requirement that the electron orbitals be constrained by 
the symmetry of the external confining field [thus the designation ``space 
(S) unrestricted''].

Subsequent to our discussion of symmetry breaking and formation of electron
molecules in the framework of the sS-UHF method, we elaborate in section IV
on the connection between the symmetry-broken sS-UHF solution and the exact
one in the case of a clean 2e QD with $B=0$. In particular, analysis of the 
exact solution using the conditional-probability-distribution (CPD) technique
\cite{berr} reveals the formation of an electron molecule in agreement with 
the sS-UHF result.

\section{Thomson quantum dot with outside impurities}

To introduce some of the principal physical and methodological issues
pertaining to symmetry breaking and formation of Wigner molecules in finite 
fermion systems, we discuss first the 2$e$ TQD at $B=0$ [we remind the reader 
that a sufficiently high value of $R_W$ ($=2.39$) was chosen,
such that the QD is in the regime where the two electrons form a WM; for the
other parameters used throughout this paper, see Ref.\ \onlinecite{note11}].
The single-particle wave functions (modulus square) and total electron 
densities displayed in Fig. 1 are taken from calculations for the 2$e$ TQD in 
the presence of two attractive impurities (${\cal Q}_1={\cal Q}_2=-1e$,
represented in the figure by filled dots on the two sides of the QD)
located symmetrically about the center of the QD at $(x,y,z)$=
$(\pm 60,0,10)$ nm with the strength and location of the impurities 
purposefully chosen such that they will not affect the nature of the 
electronic ground states (for the same QD but without the impurities), except 
for orientational pinning in the case of symmetry-broken states (see below).

Constraining first the solution to
maintain the symmetry of the hamiltonian, in conjunction
with double-occupancy of the HF orbitals by electrons of opposite spins
[that is through the use of the RHF method with input trial electron densities
satisfying the symmetry of the external potentials \cite{so}], 
the resulting symmetry-adapted (SA) self-consistent 
singlet (${\cal S}$) orbitals and corresponding total density 
distributions exhibit, as expected, an almost circular symmetry with 
minimal elliptical distortions (due to the impurities), see Fig.\ 1(a);
without the impurities, the SA singlet is perfectly circularly symmetric.

However, increasing the variational
freedom through removal of the spatial symmetry and double-occupancy 
constraints via the use of the sS-UHF method results in a symmetry-broken
singlet ground state of lower energy, that is formation of a WM characterized 
by localized orbitals with the ``bond length'' (distance between the maxima 
in the total electronic distribution) equal to 29 nm [see Fig. 1(b)]; 
the energy of this state is lower by 1.62 meV than that of the SA solution
[Fig.\ 1(a)]. This lowering of the ground-state energy reflects gain in 
correlation energy (for the definition of correlation energy, see section
IV below). Note that the 2e WM is orientationally pinned along the 
interimpurity axis. Similar formation of a 
WM occurs also for a $2e$ TQD without the pinning
impurities (with an energy gain of 1.32 meV compared to the corresponding SA 
solution). However, in the absence of pinning, the formation of the WM is 
accompanied by orientational degeneracy \cite{note222} (that is there is an 
infinite manifold of rotationally degenerate sS-UHF ground states). 

The formation of a fermionic molecule, associated with electron localization,
in the ground-state of a QD under magnetic-field-free conditions does not have
an analog within the framework of the traditional models of atomic structure.
\cite{cs} However, the physics underlying this phenomenon, which is a
manifestation of SB resulting from strong electronic correlations (see
also section IV below), resembles closely that found in doubly-excited 
two-electron atoms. Indeed, spectroscopical studies on doubly-excited
helium atoms revealed rovibrational bands which were interpreted, borrowing
from models developed in the context of nuclear and molecular physics,
by invoking the formation of a ``triatomic'' molecule comprised of the two
localized electrons and the He$^{2+}$ nucleus ($\alpha$ particle), with the
collinear configuration being of particular significance. \cite{note111}

The first electronically excited state of the $2e$ TQD is the triplet 
${\cal T}$
state (with the spins of the two electrons parallel to each other) whose
total electron density distribution [Fig.\ 1(c), left] resembles that of the 
ground-state singlet [Fig.\ 1(b), left]. However, the individual electronic
wave functions in the ${\cal T}$ state differ in character from those of the
${\cal S}$ state [compare right panels in Fig.\ 1(c) and 1(b)], 
with the lower-energy 
one being $s$-like (but elliptically distorted), and the other is a $p$-like
orbital oriented by the impurities along the $x$-axis. Note that the ${\cal T}$
state has the symmetry of the hamiltonian including the two pinning impurities
(that is, here the sS-UHF solution coincides with the symmetry-adapted one).
The same orbital characters are obtained also in the absence of the pinning 
impurities, but without a preferred orientation. In this case, however,
the character of the ${\cal T}$ state is a result of spontaneous SB, with an 
energy gain of 0.09 meV compared to the corresponding SA (circular) solution.
Underlying the type of spontaneous SB in the (open-shell) ${\cal T}$ state of
the $2e$ TQD is the Jahn-Teller (JT) effect \cite{jt} where lowering of the
total energy is achieved via mixing of the two-fold degenerate $m=+1$
($p_+$) amd $m=-1$ ($p_-$) angular momentum states, concomitant 
with a deformation of the self-consistent potential away from circular 
symmetry. \cite{note1} To distinguish such electron molecules from the WM
discussed above for the closed-shell singlet state (whose formation is
driven by the dominance of the electron-electron repulsion), we refer to them
as JT electron molecules (JTEM's). Such spontaneous SB via JT distortions 
is familiar from studies of the rotational spectra of open-shell 
nuclei \cite{bm,naza} and from investigations of shape deformations of 
open-shell metal clusters. \cite{yl1,ekar} 

Similar calculations for the $2e$ TQD, 
but with repulsive pinning impurities (that is ${\cal Q}_1={\cal Q}_2=+1e$) 
yield for the singlet ground state
qualitatively similar results (with different values for the energies), but
with an important distinction that now the pinned orientation of the WM is 
rotated by $\pi/2$ compared to the ${\cal Q}_1={\cal Q}_2=-1e$ case
(i.e., the ``intramolecular'' axis of the WM is oriented normal to the
interimpurity axis). Consequently, through variation of the sign (polarity) of
the impurity gate voltages, one may ``flip'' the orientation of the WM, and 
with it the direction of the polarization of the electronic charge 
distribution in the QD. 
In this way, the formation of WM's in QD's and the ability to control their 
discrete orientations via pinning voltage gates may serve as a method 
for the creation of on-off information storage cells and nanoscale logic
gates. \cite{lent}

\section{Rutherford quantum dot}

Next, we examine the properties of a $2e$ Rutherford QD (RQD), that is a
$2e$ QD with a central attractive impurity. The sS-UHF singlet and triplet
electronic orbitals corresponding to a $2e$ RQD for $B=0$ with a single
impurity (${\cal Q}=-2e$) located at (0,0,10) nm are shown in Fig.\ 
2(b) and 2(c) respectively. They exhibit WM symmetry breaking and
JT-distortion features similar to those found for the $2e$ TQD [compare
Fig.\ 1(b and c)], but with a reduced WM bond length and a more compact
triplet. The ``strength'' of the SB depends of course on the impurity charge
${\cal Q}$ and/or its distance $d$ from the QD plane. For example, for
${\cal Q} = -1e$ and $d=0$ (and for an arbitary position of the
impurity inside the QD), no symmetry breaking was found by us due to the
strong trapping by the impurity of the two electrons which occupy
circularly symmetric orbitals, resembling the behavior of the 
ground-state of the natural He atom. \cite{note33} We also remark that for 
the case described in Fig.\ 2(b) and Fig.\ 2(c) both electrons are slightly 
trapped by the impurity potential for the ${\cal S}$ state, while for the 
${\cal T}$ state the $s$-like electron is strongly trapped and the $p$-like 
electron occupies a much less bound orbital.

Note that here, as with the case of a clean QD, the singlet WM and the 
JT-distorted triplet are free to rotate in the plane of the QD, since there 
are no off-centered pinning impurities. 

The large physical size of QD's makes them 
ideally suited for investigations of magnetic-field effects and controlled 
manipulations. To illustrate such effects, we display in Fig.\ 2(a) 
the magnetic-field induced variation of the total energies 
of the ${\cal S}$ and ${\cal T}$ states in the $2e$ RQD (qualitatively
similar behavior is found also for the $2e$ TQD).
As seen, the energy of the singlet
state increases and that of the triplet state decreases with increasing
$B$; for fields $B < B_1$ (0.2 T), the variation of the energy of the 
${\cal T}$ (the slope of the curve) is smaller than that for $B > B_1$
[see inset in Fig.\ 2(a)]. Furthermore, at a critical value
$B_c=2.8$ T [marked by a down-arrow in Fig.\ 2(a)], the energies of the two 
states cross and from then on the triplet lies below the singlet. 

For the singlet, the broken-symmetry WM state 
[Fig.\ 2(b)] maintains under the influence of the
applied magnetic field in the range considered in Fig.\ 2(a), 
with the increase in the magnetic-field strength leading to further
shrinkage of the bond length of the WM accompanied by an overall increase of 
the energy of the ${\cal S}$ state \cite{note2} [see Fig.\ 2(a)].

The influence of the magnetic field on the triplet state is more subtle.
As aforementioned, at $B=0$ the symmetry of the ${\cal T}$ state is broken by 
the JT effect involving mixing of the $m=\pm 1$ degenerate angular momentum 
orbitals (see Ref.\ \onlinecite{note1}).
On the other hand, the magnetic field lifts the degeneracy of these $p_+$ and
$p_-$ states (without mixing them), and this effect competes with the 
JT distortion. For small enough fields [$B < B_1$, see inset in Fig.\ 2(a)], 
the JT effect prevails, and thus the orbitals and electron densities 
remain similar to those shown in Fig.\ 2(c), and they maintain an 
orientational degeneracy in the plane.
At stronger fields ($B > B_1$), the lifting of the energetic degeneracy of the
$p_+$ and $p_-$ states overcomes the JT 
effect, and the second electron populates the lower of these two orbitals.
As a result, circular symmetry is recovered, as illustrated in Fig.\ 2(d).

The overall decrease with $B$ of the energy of the ${\cal T}$ state relative
to the ${\cal S}$ state is due to enhanced stabilization by the 
(parallel-spin) exchange energy in the former, reduced Coulomb repulsion 
between the electrons occupying $s$- and $p$-like orbitals, and quenching
of the kinetic energy of the $p$-like orbital by the magnetic field.
\cite{note2} This effect increases with $B$, and at $B \geq B_c$ the 
${\cal T}$ state becomes the ground state [see Fig.\ 2(a)].
Note that this transition is driven primarily by the interelectron 
repulsion and not by the interaction of the electrons' moments with 
the magnetic field (see Ref.\ \onlinecite{asho} and references therein);
for our system, the Zeeman splitting energy is 0.026 meV/T.
At even larger fields, the Coulomb repulsion between the electrons
increases (due to the shrinking of the orbitals) resulting in an ascending
trend of the energy of the ${\cal T}$ state, which remains, however, lower
than the singlet state. A similar scenario is found also for the TQD
(without pinning impurities).

In light of previous findings \cite{maks,mk} pertaining to formation of fully
spin-polarized symmetry-broken states in clean QD's (TQD's) at high magnetic
fields, it is pertinent to inquire whether the circular symmetry found for
the ${\cal T}$ state of the $2e$ RQD for $B_1 < B < 5$ T will also be
broken at higher fields. For the clean $2e$ TQD, we verified that
indeed an orientationally degenerate electron molecule [with the molecular 
orbitals of the electrons distributed about the two molecular centers, see 
Fig.\ 2(e)] formed at sufficiently high $B$ (e.g., $B=10$ T). Such an electron
molecule is akin \cite{note3} to the JTEM discussed above in the context of the
triplet state for $B=0$. Interestingly, such reemergence of a JTEM structure 
does not occur at these conditions for the $2e$ RQD studied here due to the 
enhanced gap between the $p_+$ orbital and the strongly trapped $s$ orbital.
This provides an additional venue for impurity-assisted manipulation and 
design of the electronic properties of QD's.

\section{Connection to the exact solution}

As mentioned earlier, for $R_W > 1$, the sS-UHF approach applied to QD's 
yields {\it approximate\/} ground-state solutions which violate the symmetries
of the original many-body hamiltonian, e.g, the spontaneous 
breaking of rotational symmetry discussed in sections II and III for a
circularly symmetric clean QD (i.e., a TQD) or one with a central impurity
(i.e., RQD). At a first glance, this situation may appear puzzling, but it is 
not unique in the context of many-body theory of finite fermionic systems. 
Indeed, a similar situation was encountered in nuclear physics in the 
1950's, when it was discovered that open shell nuclei carried
permanent quadrupole moments and that many of them exhibited well developed
rotational spectra (i.e., they behaved like rigid rotors). The explanation of 
these experimental findings was formulated in the framework of breaking of the
rotational symmetry associated with nuclear deformations of the Jahn-Teller 
type, and it led to several celebrated models and semi-empirical methods, 
i.e., the particle-plus-rotor model of Bohr and Mottelson, \cite{bm2} the 
modified-anisotropic-oscillator model of Nilsson 
\cite{nils} and the Strutinsky shell-correction method. \cite{stru} 
In the language used by us here, this means that at the microscopic level 
the breaking of the rotational symmetry had to be accounted \cite{ripk}
for via space (S)-UHF methods (that is allowing the spatial-orbital solutions
of the Hartree-Fock equations to assume symmetries lower than those of
the underlying many-body hamiltonian).

Starting with Peierls and Yoccoz \cite{py} (see also Peierls and Thouless
\cite{pt}), numerous theoretical investigationss have addressed 
the connection of the broken-symmetry HF solution to the exact solution, and
led to the theory of {\it restoration-of-symmetry\/} via Projection
Techniques. \cite{proj}
The central physical ingredient of the Projection Technique rests
with the observation that a HF solution which breaks rotational symmetry is not
unique, but belongs to an infinitely-degenerate manifold of states with
different spatial orientation. A proper linear combination of the HF 
determinants in such a manifold yields multi-determinental states 
with good {\it total\/} angular momenta that are a better approximation to the 
exact solution. For a comprehensive review on restoration of symmetry in the
context of nuclear many-body theory, we refer the reader to the book by
Ring and Schuck mentioned in Ref.\ \onlinecite{note222} (see also Ref.\
\onlinecite{bro}, where the principles of restoration of rotational symmetry 
are discussed in the two-dimensional case).

The restoration of symmetry via Projection Techniques in the case of  
sS-UHF solutions describing 2D-QD's will be presented in a future publication.
In this section, we found it more convenient to discuss the connection between 
the sS-UHF and the exact solutions by taking advantage of the simplicity of 
solving the exact problem at $B=0$ for two electrons interacting via the 
Coulomb force and confined by an external parabolic confinement without 
impurities (clean QD). Indeed, it is well known that in this case, the exact 
Schr\"{o}dinger equation for two interacting electrons is separable in the
center-of-mass, ${\bf R}=({\bf r}_1
+ {\bf r}_2)/2$ (with a corresponding mass ${\cal M}=2m^*$), and relative, 
${\bf r}={\bf r}_1 - {\bf r}_2$ (with a corresponding reduced mass 
$\mu=m^*/2$), coordinates. 

To analyze the properties of the exact solutions, two quantities 
\cite{pfann,taut} have customarily been extracted from the two-body wave 
function $\Psi(\widetilde{\bf r}_1,\widetilde{\bf r}_2)$ [where the tilde
denotes both spatial and spin variables, i.e.,
$\widetilde{\bf r}_i \equiv ({\bf r}_i, s_i)$ $(i=1,2)$, see Appendix 
B]: (i) the usual pair-correlation function, \cite{note23}
\begin{equation}
G(v)= 2\pi \langle \Psi | \delta({\bf r}_1-{\bf r}_2-{\bf v}) | \Psi \rangle~,
\label{pair}
\end{equation}
and (ii) the electron density (ED)
\begin{equation}
n(v)=\langle \Psi |\sum_{i=1}^2 \delta({\bf v}-{\bf r}_i) |\Psi \rangle~.
\label{spdens}
\end{equation}

However, for the exact $\Psi$ in the case of a circularly symmetric 
confinement, both of these quantities turn out to be also
circularly symmetric and thus they do not reveal 
the full physical picture, the (possible) formation of an electron molecule
generated by electron correlations. A more suitable quantity here is the 
conditional probability distribution (CPD) 
${\cal P}({\bf v}|{\bf r}_2={\bf v}_0)$ for 
finding one electron at ${\bf v}$ given that the second electron is at 
${\bf r}_2={\bf v}_0$. This quantity has been extensively used \cite{berr} in 
the analysis of electron correlations in doubly-excited helium-like atoms, and
is defined as follows,
\begin{equation}
{\cal P}({\bf v}|{\bf r}_2={\bf v}_0) \equiv
\frac{ \langle \Psi | \delta({\bf v}-{\bf r}_1) \delta({\bf v}_0-{\bf r}_2) 
| \Psi \rangle } { \langle \Psi | \delta({\bf v}_0-{\bf r}_2) | \Psi \rangle}~
\label{cpd}
\end{equation}
[for details concerning the calculation of the quantities in Eqs.\
(\ref{pair}) $-$ (\ref{cpd}), see Appendix B].

In the left column of Fig.\ 3, we display the above three quantities for the
exact ground state in the case of a parabolic confinement with the same 
parameters as used throughout this paper (see Ref.\ \onlinecite{note11}) and
$B=0$. The pair-correlation function in Fig.\ 3(a) exhibits a well developed 
depression at $v=0$, namely the two electrons on the average keep apart from 
each other at a distance $2r_0=22.92$ nm. In Fig.\ 3(b), we plot the CPD with 
${\bf v}_0=(r_0,0)$ (marked by a cross). It is seen that the maximum 
probability for finding the second electron occurs at the diametrically 
opposite point $-{\bf v}_0$, in accordance with the picture 
of an electron molecule presented earlier within the sS-UHF approach.
The exact electron density shown in Fig.\ 3(c) is circularly symmetric, as
expected. Comparing figures 3(b) and 3(c), the following 
interpretation ensues naturally, namely that the CPD in Fig.\ 3(b) describes 
the electron molecule in its {\it intrinsic (body-fixed)} frame of reference, 
while the electron density in Fig.\ 3(c) describes the electron molecule in 
the laboratory frame of reference where rotational and and center-of-mass 
displacements are superimposed upon the intrinsic probability density.

Fig.\ 3(d) displays the electron density for the corresponding sS-UHF
ground state. As discussed earlier, the sS-UHF electron density breaks the 
rotational symmetry and clearly exhibits the morphology of an electron 
molecule, unlike the exact one in Fig.\ 3(c). It is apparent that the sS-UHF 
electron density corresponds to that in the intrinsic frame of the electron 
molecule. Restoration of the symmetry via Projection Techniques will bring
the sS-UHF electron density closer to that of the exact solution. As
aforementioned, this interpretation is familiar in the context of nuclear 
and it is further supported by the CPD calculated with the sS-UHF ground-state
[i.e., by using $\Psi^{\text{UHF}}$ in Eq.\ (\ref{cpd}), instead of the exact 
many-body $\Psi$, see Appendix B] and displayed in Fig.\
3(e). Although naturally not identical, the two CPD's (i.e., the exact and the
UHF) are of similar nature and both illustrate 
graphically the correlation effect associated with electron localization and
formation of an electron molecule. We further illustrate this point by
contrasting the exact and sS-UHF CPD's with the CPD of the symmetry-adapted
RHF ground state, shown in Fig.\ 3(f). In this latter case,
$\Psi^{\text{RHF}}=\psi^{\text{RHF}}({\bf r}_1) 
\psi^{\text{RHF}}({\bf r}_2) \chi(s_1,s_2;S=0)$, with 
$\psi^{\text{RHF}}({\bf v})$ being the 1$s$ orbital of the RHF; as 
a result, the probability of finding one electron at ${\bf v}$ is independent
of the position of the second electron and it is centered about the origin
where it achieves its maximum value, as expected from an independent-particle 
description, 
i.e.,
\begin{equation}
{\cal P}^{\text{RHF}}({\bf v}|{\bf r}_2={\bf v}_0) =
|\psi^{\text{RHF}}({\bf v})|^2~.
\label{cpdrhf}
\end{equation}

Finally it is of interest to examine the energetic aspects of the symmetry
breaking. Indeed the energy of the ground-state is 
$E^{\text{RHF}}_{\text{gs}}=22.77$ meV for the RHF solution, 
$E^{\text{UHF}}_{\text{gs}}=21.45$ meV for the sS-UHF solution, and 
$E^{\text{ex}}_{\text{gs}}=19.80$ meV for the exact one. 
Since the correlation
energy is the difference between the RHF and the exact energies, \cite{cor}
one finds $\varepsilon^{\text{corr}}_{\text{gs}}=E^{\text{RHF}}_{\text{gs}}-
E^{\text{ex}}_{\text{gs}}=2.97$ meV. Thus the symmetry breaking associated 
with the sS-UHF solution is able to capture 
$(E^{\text{RHF}}_{\text{gs}}-E^{\text{UHF}}_{\text{gs}})/
\varepsilon^{\text{corr}}_{\text{gs}}=$ 44\% of the correlation energy; the 
remaining amount can be captured through improvements 
via Projection techniques.
\cite{yl3}

\section{Summary}

The dominance of interelectron interactions and correlations in quantum
dots (often referred to as ``artificial atoms'') results in spontaneous
symmetry breaking of the self-consistent mean field, beyond the central-field 
picture on which the modern understanding of atomic structure has been founded.
Indeed, as shown in this paper through spin-and-space unrestricted 
Hartree-Fock \cite{yl2,note12} many-body calculations for the singlet
ground state of a 2e QD, such symmetry 
breaking and the resulting formation of Wigner molecules do occur for a wide 
range of system parameters (with or without impurities) even in the most 
elementary case of two-electron QD's. 

Additionally, in the case of fully polarized dots (e.g., the triplet state of 
a 2e QD), the sS-UHF calculations show that similar electron molecules form
due to symmetry breaking  associated with Jahn-Teller-type distortions.

In spite of the simplicity of the hamiltonian, correlations in two-electron
QD's underlie a remarkably rich and complex physical behavior. As a result,
2e QD's can serve as paradigms for a unification of concepts pertaining to 
spontaneous SB in various finite fermion aggregates, including nuclei, excited
atoms, and clusters. 

Furthermore, impurity and/or magnetic-field effects allow for 
controlled manipulation of the formation and pinning of the discrete 
orientations of the electron molecules in 2e QD's. Such ability may be 
employed in future applications of QD's as nanoscale logic cells and 
information storage elements. Impurity effects were also illustrated for the 
case of a quantum dot with more than two electrons.

Confirmation of the formation in 2e QD's of electron molecules associated
with symmetry breaking of the self-consistent field (in the context of 
sS-UHF calculations) was obtained through an analysis of the exact ground
state via the conditional-probability-distribution technique.

\acknowledgments

This research is supported by the US D.O.E. (Grant No. FG05-86ER-45234).

\appendix

\section{QD's with more than two electrons}

In the main body of this paper we limited ourselves to the case of
2e QD's, since focusing on this elementary case allowed us to better elucidate
the intricate physical principles involved in the formation and impurity 
control of electron molecules in QD's, from both the perspective of the sS-UHF
treatment and the exact solution. In this appendix, we will present an example
of the many different structural possibilities that can arise when impurities 
are introduced in a QD with a larger number of electrons. In particular, we 
consider a QD with 8 electrons in the presence of a hydrogenic-like $(d=0)$ 
impurity of variable nominal charge ${\cal Q}=qe$ placed at the center (and in
one case off-center) of the QD (for the other input parameters, we use same 
values as used throughout the text, see Ref.\ \onlinecite{note11}). 
Since this appendix does not intend to present an exhaustive study 
of larger dots, but simply aims at presenting an illustrative example,
we will consider only one spin configuration, i.e., the sS-UHF solutions 
having 4 spin-up and 4 spin-down electrons. However, $q$ will be allowed 
to take both positive (repulsive) and negative (attractive) values.

Fig.\ 4(a) displays the sS-UHF electron density for a repulsive central
impurity [located at (0,0,0)] with ${\cal Q}=+1e$. It is seen that a Wigner 
molecule consisting of a single ring with 8 electrons [denoted as a (0,8) 
ring] is formed in this case. For a slightly attractive central impurity with 
${\cal Q}=-0.2e$, however, a structural change takes place, namely one 
electron moves to the center of the dot [see Fig.\ 4(b)]. Notice that this 
(1,7) structure agrees with the arrangement found in studies of classical 
point-charges in a purely $({\cal Q}=0)$ harmonic confinement. \cite{lp} 
Increasing the attractive nominal charge to ${\cal Q}=-0.5e$ [see Fig.\ 4(c)] 
results in a further structural change, namely the central impurity is now 
able to trap two electrons, thus leading to a (2,6) arrangement. A further 
increase of the attractive charge of the impurity to the value ${\cal Q}=-1e$ 
does not produce any qualitative change in the (2,6) arrangement as long as 
the impurity remains at the center of the dot. By moving the impurity to an 
off-center position, however, various structural morphologies 
can arise, an example of which is presented in Fig.\ 4(d) for a ${\cal Q}=-1e$ 
impurity located at (20,0,0) nm, forming a highly distorted (2,6) WM
(here the first index denotes that two electrons are trapped by the impurity).

A further increase in the attractive charge ${\cal Q}$ leads to sequential
trapping of the remaining six electrons and to progressive elimination of
symmetry breaking, until all eight electrons have been captured by the
impurity (see also section III).

\section{Two-electron wave functions and the definitions (3)-(5)}

\subsection{Exact solution for two electrons}

In the case of two interacting electrons confined by a parabolic potential
(clean QD), one can perform a change of variables to center-of-mass (CM),
${\bf R}=({\bf r}_1+{\bf r}_2)/2$ and ${\bf P}={\bf p}_1+{\bf p}_2$, and
relative-motion (rm), ${\bf r}={\bf r}_1-{\bf r}_2$ and ${\bf p}=({\bf p}_1-
{\bf p}_2)/2$, coordinates. Then the two-electron hamiltonian separates into 
CM and rm contributions, 
\begin{equation}
{\cal H}={\cal H}_{\text{CM}}+ {\cal H}_{\text{rm}}~,
\label{h12}
\end{equation}
with
\begin{equation}
{\cal H}_{\text{CM}} =
\frac{{\bf P}^2}{2 {\cal M}} + \frac{1}{2} {\cal M} \omega_0^2 R^2~,
\label{hcm}
\end{equation}
and
\begin{equation}
{\cal H}_{\text{rm}}= 
\frac{{\bf p}^2}{2 \mu} + \frac{1}{2} {\mu} \omega_0^2 r^2 +
\frac{e^2}{\kappa r}~,
\label{hrm}
\end{equation}
where ${\cal M}=2m^*$, $R=|{\bf R}|$, $\mu = m^*/2$, and $r=|{\bf r}|$.

The center-of-mass motion associated with the coordinate ${\bf R}$ obeys a 
Schr\"{o}dinger equation describing the motion of a particle of mass 
${\cal M}=2m^*$ in a 2D isotropic harmonic potential of frequency $\omega_0$.
Here $\omega_0$ is the frequency of the original parabolic confinement, i.e.,
the interelectron repulsion has no baring on the center-of-mass motion.

Using dimensionless polar coordinates $U=R/(l_0 \sqrt{2})$ and 
$\Theta$, the center-of-mass wave function can be written as
$\Xi (U) e^{iM\Theta}$ with the radial part given by
\begin{equation}
\Xi(U)=C_{NM} U^{|M|} e^{-U^2/2{\cal L}_0^2} L^{|M|}_N(U^2/{\cal L}_0^2)~,
\label{cmwf}
\end{equation}
where $(N,M)$ are the radial and azimuthal (related to the angular momentum) 
quantum numbers, respectively, ${\cal L}_0=1/2$, the normalization constant 
$C_{NM}=[2 N! 4^{|M|+1}/(N+|M|)!]^{1/2}$, 
and $L^{|M|}_N(x)$ are associated Laguerre polynomials.

Since in the exact problem the Coulomb interaction preserves the rotational 
symmetry, the radial part of the wave function 
$\Omega(u) e^{im\theta}/\sqrt{u}$ 
associated with the relative motion obeys the following one-dimensional 
Schr\"{o}dinger equation [in dimensionless polar coordinates 
$u=r/(l_0\sqrt{2})$ and $\theta$],
\begin{equation}
\frac{\partial^2 \Omega}{\partial u^2} + 
\{ \frac{-m^2+1/4}{u^2}-u^2-\frac{R_W \sqrt{2}}{u}+
\frac{\varepsilon}{\hbar \omega_0 /2} \} \Omega =0~.
\label{relmot}
\end{equation}
The $1/u$ term results from the interelectron repulsion.

Defining $\widetilde{\bf u}_i \equiv ({\bf u}_i, s_i)$ (where $s_i$ is the
spin of the $i$th electron and $i=1,2$), 
the exact many-body (here two-body) wave function is given by
\begin{eqnarray}
\Psi(\widetilde{\bf u}_1,\widetilde{\bf u}_2)&=&
\Phi({\bf u}_1, {\bf u}_2) \chi(s_1,s_2) \nonumber \\
&=& \frac{1}{2\pi} \frac{\Omega(u)}{\sqrt{u}} e^{im\theta}
\Xi(U) e^{iM\Theta} \chi(s_1,s_2)~,
\label{exwf}
\end{eqnarray}
with ${\bf u}_i={\bf r}_i/(l_0 \sqrt{2})$ and $\chi(s_1,s_2)$ is the
spin part.

The exact eigenvalues are given by
\begin{equation}
E_{NM,nm}=\hbar \omega_0(2N+|M|+1)+\varepsilon(n,|m|)~,
\label{energy}
\end{equation}
where $\varepsilon(n,|m|)$ are the eigenvalues associated with the relative
motion [see Eq.\ (\ref{relmot})], $(n,m)$ being the corresponding radial and
azimuthal quantum numbers.

\subsection{Pair correlation, electron density, and conditional
probability distribution for the exact and sS-UHF two-electron wave functions}

The bracket notation in Eqs.\ (\ref{pair}), (\ref{spdens}), and (\ref{cpd})
imply integrations over both the spatial and spin variables. In the case of
the exact wave function $\Psi$ given by Eq.\ (\ref{exwf}), the spin variables 
separate out, and thus $G(v)$, $n(v)$, and the CPD can be expressed as 
double integrals over the positions $({\bf r}_1,{\bf r}_2)$ of the following 
two-body spatial probability density $W({\bf r}_1,{\bf r}_2)
=|\Phi({\bf r}_1,{\bf r}_2)|^2$, where $\Phi$ is the spatial part of $\Psi$.
In particular, one finds
\begin{equation}
G(v)=2\pi \int \int \delta({\bf r}_1-{\bf r}_2-{\bf v}) 
W({\bf r}_1,{\bf r}_2) d{\bf r}_1 d{\bf r}_2~,
\label{pair2}
\end{equation}
for the pair-correlation function,
\begin{equation}
n(v)=\int \int \sum_{i=1}^2 \delta({\bf v}-{\bf r}_i) 
W({\bf r}_1,{\bf r}_2) d{\bf r}_1 d{\bf r}_2~,
\label{spdens2}
\end{equation}
for the electron density, and
\begin{equation}
{\cal P}({\bf v}|{\bf r}_2={\bf v}_0)=
\frac{W({\bf v},{\bf r}_2={\bf v}_0)}
{\int d{\bf r}_1 W({\bf r}_1,{\bf r}_2={\bf v}_0)}~.
\label{cpd2}
\end{equation}
for the conditional probability distribution.

In the case of the sS-UHF singlet ground state, the two electrons occupy
two different spatial orbitals $\psi_1({\bf r})$ and $\psi_2({\bf r})$. Then
the corresponding two-body wave function is the following single determinant,
\begin{eqnarray}
\Psi^{\text{UHF}} ({\bf r}_1, {\bf r}_2) =
\frac{1}{\sqrt{2}} &[&\psi_1({\bf r}_1)\alpha(1)\psi_2({\bf r}_2)\beta(2)
\nonumber \\
&& - \psi_1({\bf r}_2)\alpha(2)\psi_2({\bf r}_1)\beta(1)]~,
\label{uhfwf}
\end{eqnarray}
where $\alpha$ and $\beta$ denote the spin-up and spin-down spinors,
respectively. Integration of $|\Psi^{\text{UHF}}|^2$ over the spin variables 
yields the following two-body spatial probability density,
\begin{equation}
W^{\text{UHF}}({\bf r}_1,{\bf r}_2)=
\frac{1}{2}[|\psi_1({\bf r}_1)|^2|\psi_2({\bf r}_2)|^2+
|\psi_1({\bf r}_2)|^2|\psi_2({\bf r}_1)|^2]~.                 
\label{uhfw}
\end{equation}

To calculate the Conditional Probability Distribution in the case of the 
sS-UHF ground state, one replaces $W$ by $W^{\text{UHF}}$ in Eq.\ (\ref{cpd2}).

\begin{figure}
\caption{
Total electron densities (left frame in each pannel) and contours of modulus
square of the individual orbitals (right frames) for the clean (Thomson)
$2e$ TQD at $B=0$, in the presence of two attractive (${\cal Q}_1=
{\cal Q}_2=-1e$) impurities whose projected positions on the QD plane
($xy$) are denoted by filled circles.
(a) The symmetry-adapted (RHF) singlet (${\cal S}$).
The energy of the SA singlet is {\it higher\/} by 1.34 meV than that of the
corresponding triplet, manifesting a shortcoming of the RHF method.
(b) The sS-UHF Wigner-molecule singlet, orientationally pinned along the 
inter-impurity axis and exhibiting localized orbitals, one to the left and the
other to the right of the QD middle; note the lack of reflection
symmetry of the individual orbitals about the mirror ($yz$) plane normal to the
inter-impurity axis and passing through the QD's center.
When the WM singlet is rotated by $\pi/2$, the energy of the system rises
by 0.58 meV. (c) The triplet (${\cal T}$) state 
with an elliptically deformed $s$-like and a $p_x$-like
orbital, orientationally pinned by the impurities. The energy of the 
symmetry-broken singlet in (b) is {\it lower\/} by 0.28 meV than that 
of the triplet in (c), compared to 0.43 meV for the TQD without pinning 
impurities. Distances are in nm and the electron densities in 
10$^{-4}$ nm$^{-2}$. The parameters characterizing the QD shown here are 
those given in Ref.\ \protect\onlinecite{note11}.
}
\end{figure}

\begin{figure}
\caption{
(a-d) Energetics and individual orbitals (modulus square)
from sS-UHF calculations for the $2e$ (Rutherford) RQD, with a central impurity
${\cal Q}=-2e$ at (0,0,10) nm. Energies (in meV) of the ${\cal S}$ and 
${\cal T}$ states versus $B$ (in Tesla), crossing at $B_c=$ 2.8 T (marked by
an arrow), are shown in (a). An expanded view of the energy of the ${\cal T}$ 
state for small fields, exhibiting a transition from the spontaneous JT regime
to a circular symmetric state at $B > B_1$ ($B_1$= 0.2 T),
is shown in the inset; included also is the energy of the SA solution
(dashed curve). At $B=0$, contours of the orbitals of the two electrons for the
WM singlet are shown (superimposed on each other) in (b), with one of the
orbitals depicted by a solid line and the other by a dashed line. The  
orbitals for the spontaneously JT-distorted triplet at $B=0$ are shown in the 
two panels of (c). For the singlet, the energy gain due to SB (that is 
lowering of the total energy with respect to the symmetry-adapted RHF state) 
is 0.39 meV and the energy gain due to the JT-distortion of the triplet is 
0.07 meV; the energy difference between the ${\cal T}$ and ${\cal S}$ states 
is 2.01 meV. The transition to a circular symmetric triplet 
is illustrated for $B=0.6$ T in (d). 
(e) Individual orbitals for the symmetry-broken ground-state triplet of the 
clean (Thomson) $2e$ TQD at $B=10$ T. Under the same conditions, the ground
state triplet in the $2e$ RQD remains circularly symmetric.
Distances in nm and orbital densities in 10$^{-4}$ nm$^{-2}$.
Note the different length scales of the $(x,y)$ axes in (b-e) compared to
those in Fig.\ 1. The parameters characterizing the QD's shown here are 
those given in Ref.\ \protect\onlinecite{note11}.
}
\end{figure}

\begin{figure}
\caption{
Ground-state results for a clean 2e QD with the parameters given in Ref.\
\protect\onlinecite{note11} and $B=0$; in all cases the ground state is a 
singlet. (a-c) correspond to the exact solution. (a) The
pair correlation function [$G(v)$, see Eq.\ (\protect\ref{pair})] plotted 
versus $v$ exhibiting a well developed depression at $v=0$, with a mean 
electron-to-electron separation of $2r_0=22.92$ nm. (b) The electron 
conditional probability distribution [CPD, see Eq.\ (\protect\ref{cpd})] with 
${\bf v}_0=(r_0,0)$ (denoted by a cross), showing formation of a 2e electron 
molecule. (c) The electron density [ED, see Eq.\ (\protect\ref{spdens})], 
reflecting the conservation of circular symmetry by the exact two-electron 
solution. The results displayed in (d) and (e) correspond to calculations 
using the sS-UHF method. Spontaneous symmetry breaking leading to formation of
a 2e-molecule (with a bond length of $2r_0^\prime=28.16$ nm) is exhibited by 
the sS-UHF electron density shown in (d). (e) Such an electron molecule is 
further reflected in the sS-UHF CPD with ${\bf v}_0=(r_0^\prime,0)$ (denoted 
by a cross). (f) The CPD with ${\bf v}_0=(r_0^\prime,0)$ 
(denoted by a cross) corresponding to the restricted Hartree-Fock
(RHF) ground state (i.e., the symmetry-adapted solution) exhibits, as
expected, no symmetry-breaking signature. Lengths are in nm and 
density functions [$G(v)$, ED's, and CPD's] are in units of 10$^{-4}$
nm$^{-2}$.
}
\end{figure}

\begin{figure}
\caption{
Electron densities obtained via sS-UHF calculations of a QD containing 8
electrons at $B=0$ in the presence of a central [located at (0,0,0) in (a-c)] 
and an off-centerd [located at (20,0,0) nm in (d)] hydrogenic impurity.
The parameters characterizing the QD shown here are those given in
Ref.\ \protect\onlinecite{note11} and the charge ${\cal Q}$ of the impurity is
as marked on the figure. All the cases correspond to zero total spin, i.e.,
4e with spin up and 4e with spin down. For each of the cases, we observe 
formation of a Wigner molecule, with its structure dependent on the polarity
(sign), magnitude and location of the impuriry. (a) A repulsive central 
impurity $({\cal Q}=+1e)$, resulting in an 8e WM with a ring structure and
an empty-electron central region, denoted as (0,8). (b) Slightly attractive
central impurity $({\cal Q}=-0.2e)$ leading to formation of an (1,7) two-ring 
WM. (c) A stronger attractive central impurity $({\cal Q}=-0.5e)$ associated
with a (2,6) WM. (d) An off-center attractive impurity $({\cal Q}=-1e)$,
resulting in a distorted 6e WM with two electrons trapped by the impurity.
Lengths in nm and electron densities in units of 10$^{-3}$ nm$^{-2}$.%
}
\end{figure}


\begin{references}
\bibitem{noterw}
That $R_W$ is the only relevant parameter in the formation of a WM in a clean
QD can be seen immediately from the exact Schr\"{o}dinger equation 
describing the relative motion of two interacting electrons at $B=0$ in a
parabolic confinement (Appendix B). Indeed, this equation when expressed
in dimensionless variables as in Eq.\ (\ref{relmot}) contains $R_W$ as the
sole free parameter. The corresponding center-of-mass motion is independent
of the interelectron repulsion. For how to extract information about the
Wigner molecule from the exact two-body solution, see section IV.
\bibitem{note11}
In all calculations, we used $g^*$(GaAs)$=-0.44$, $m^*$(GaAs)$=0.067m_e$,
and $\hbar \omega_0=5$ meV, with $R_W=2.39$ (that is $\kappa=8$); the same
value of $R_W$ is obtained by using $\kappa$(GaAs)$=12.9$ and
$\hbar \omega_0=1.9$ meV, yielding the similar results given here, but with 
different absolute energy values.
\bibitem{hf}
From the point of view of the terminology adopted here, the simple designation
Hartree-Fock (HF) in the literature most often refers to the RHF, in particular
in atomic physics and the physics of the homogeneous electron gas. In nuclear 
physics, however, the simple designation HF most often refers to a space 
(S)-UHF.
\bibitem{note1}
Recall that for an electron moving in a 2D circularly symmetric 
potential the wave functions are given (in polar coordinates)
by $\Psi_{n_r,m} (r,\theta) = R(r; n_r, |m|) \exp(im\theta)$ with 
corresponding energies ${\cal E} (n_r,|m|)$, where $n_r$ is the number of
radial nodes and $m$ is the angular momentum quantum number, with
$m=0, \pm 1, \pm 2,...$ In the JT framework, the second electron of the
(open-shell) triplet state of the $2e$ QD occupies an orbital formed as a
linear combination of the $n_r=0$, $m=\pm 1$ states (denoted by $p_+$ and
$p_-$), i.e., $p_x \propto p_+ +p_-$ or 
$p_y \propto p_+ - p_-$, with the one of lower
energy being selected through the self-consistent distortion of the central
mean field. As a consequence, the orbitals of both electrons in the $2e$ QD
exhibit noncircular symmetry. The energy gain due to the JT distortion
depends on the strength of the confinement  potential, and it increases for 
weaker confinements. In natural atoms, the spontaneous JT effect is fairly 
small and perturbative methods are usually adequate; see Ch. III.2 in 
Ref.\ \onlinecite{thou}.
\bibitem{note33}
This cancellation of crystallization due to a strong-attractive-impurity 
effect is illustrated in Fig.\ 4(d) where the attractive 
impurity is able to capture two of the eight electrons in the dot. The two
captured electrons form a singlet subunit and their orbitals 
are clearly circularly symmetric (see Appendix A below).
\bibitem{note2}
The effective confinement frequency $\Omega_{\text{eff}}=
(\omega_0^2+\omega_c^2/4)^{1/2}$ 
in the Darwin-Fock model of a 2D harmonically-confined electron in the
presence of a magnetic field $B$ (where $\omega_0$ is the
frequency of the harmonic potential and $\omega_c=eB/m^*c$ is the cyclotron 
frequency) increases as a function of the magnetic field.
In a magnetic field, the energy of a harmonically confined electron 
is given in the Darwin-Fock model by $E_{n_r,m}=(2n_r+|m|+1)\hbar 
\Omega_{eff}-m \hbar \omega_c/2$.
\bibitem{note3}
Such JT-like electron molecules have been called Wigner molecules in previous
publications treating fully polarized QD's (Refs.\ \onlinecite{maks} and 
\onlinecite{mk}). However, this naming is not appropriate since their 
formation is not driven by the dominance of the interelectron repulsion
(over the zero-point energy), which for large magnetic fields 
can be treated perturbatively (Refs. \onlinecite{wagn} and \onlinecite{taut1}).
\bibitem{proj}
The concept of restoration of symmetry is not restricted to spatial
symmetry, but includes also restoration of spin symmetry (see the paper
by L\"{o}wdin in Ref. \onlinecite{note222}.
\bibitem{note23}
For the purely parabolic confinement considered here, it turns out that
the pair-correlation function reduces \cite{taut} to the modulus square of the
radial part of the wave function (see Appendix B) describing the relative 
motion, i.e., $G(v)=\Omega^2(v)/v$.
\end{references}
\end{document}